\documentclass{appolb}
\usepackage{graphicx}
\begin{document}
%\title{A New Percolation Theory for Cuprates}%
\title{IS THE SUPERCONDUCTIONG STATE FOR THE CUPRATES REACHED 
THROUGH A PERCOLATION TRANSITION?%
\thanks{Presented at the Strongly Correlated Electron Systems 
Conference, Krak\'ow 2002}%
}

% Authors and Affiliations

\author{E. V. L. de Mello, E. S. Caixeiro and J. L. Gonz\'alez 
\address{Departamento de F\'{\i}sica,
Universidade Federal Fluminense, av. Litor\^ania s/n, Niter\'oi, R.J.,
24210-340, Brazil}
%\and
%the Name(s) of other Author(s)
%\address{and their affiliation}
}
\maketitle

% Abstract

\begin{abstract}
Several recent experiments have revealed that
the charge density $\rho$  in a given compound (mostly underdoped)
is intrinsic inhomogeneous with large spatial variations. Therefore
it is appropriate to define a local charge density $\rho(r)$. These
differences in the local charge concentration  yield insulator
and metallic regions, either in an intrinsic  granular or
in a stripe morphology. In the metallic region, the inhomogeneous charge
density produces spatial or local distributions of
superconducting critical
temperatures  $T_c(r)$ and zero temperature
gap $\Delta_0(r)$.
We propose that the superconducting phase in high-$T_c$ oxides is 
reached when the temperature reachs a value which superconduction
regions  with different critical temperatures percolates.
We show also that this novel approach is able to reproduce the
phase diagram for a family of cuprates and provides new insights on
several experimental features of high-$T_c$ oxides.
\end{abstract}


% The main text
  
%\section{Introduction}
The non usual properties of  high-$T_c$ superconductors
have motivated several experiments and two features
have been discovered which distinguish them from the overdoped compounds:
firstly, the appearance of a pseudogap at a temperature $T^*$,
that is, a discrete structure
of the energy spectrum above $T_c$, identified by
several different probes\cite{TS}
Second, there are increasing
evidences that the electrical charges are highly inhomogeneous
up to (and even further) the optimally
doped region\cite{Egami96,Pan,Davis}.
In fact, such intrinsic inhomogeneities  are also consistent
with the presence of charge domains either in a granular
\cite{Pan,Davis} or
in a stripe form.

In a recently letter\cite{Mello01} we have proposed a new scenario
in which a given HTSC compound with an average hole per $Cu$ ion
density  $\langle \rho \rangle$
and with an  inhomogeneous microscopic charge distribution $\rho(r)$
has a distribution of small clusters, each with a given $T_c(r)$.
This percolating scenario can be understood by analyzing
the scanning SQUID microscopy magnetic data which
makes a map of the expelled magnetic flux (Meissner effect) domains
on LSCO films\cite{IYS}. This experiment  shows the regions where the
Meissner effect continuously develops from near $T^*$ to temperatures
well below the percolating threshold $T_c$. 
Below we show that the percolating approach is capable to
yield quantitative agreement with the
HTSC phase diagrams and it
provides also new physical insight on a number of phenomena detected
in these materials.
%\section{The Charge Distribution}

Scanning tunneling
microscopy/spectroscopy (STM/S)\cite{Pan} on optimally doped
$Bi_2Sr_2CaC_2O_{8+x}$ measures nanoscale spatial variations
in the local density
of states  and the superconducting gap at a very short length
scale of $\approx 14 \AA$. These results suggest that instead
of a single value, the zero
temperature superconducting gap assumes different values at different
spatial locations in the crystal and their statistics
yield a Gaussian distribution\cite{Pan}. New high resolution STM
measurements\cite{Davis} have revealed an interesting map of
the superconducting gap spatial variation for underdoped Bi2212.

In order to model the above experimental observations
we used a phenomenological combination of a Poisson and a
Gaussian distribution for the charge distribution $\rho(r)$.Thus,
for a given compound  with an average charge density $\langle \rho \rangle $,
the hole distribution function $P(\rho; \langle \rho \rangle)$ or simply
$P(\rho)$ is a  histogram  of the  probability of
the local hole density $\rho$ inside the sample,
separated in two branches or domains.
The low density branch represents the  hole-poor or isolating regions and
the high density one represents the  hole-rich or metallic regions.
Such normalized charge probability distribution may be given by:

\begin{eqnarray}
 P(\rho) &=&  (\rho_c-\rho)\exp[-(\rho-\rho_c)^2/2(\sigma_-)^2]/ \nonumber
        [(\sigma_-)^2(2-\exp
\\ &&(-(\rho_c)^2/2(\sigma_-)^2))] \; for \; 0<\rho<\rho_c
\label{equationa}
\\
P(\rho) &=& 0  \;  for \;  \rho_c<\rho<\rho_m
\label{equationb}
\\
P(\rho) &=&  (\rho-\rho_m)\exp[-(\rho-\rho_m)^2/2(\sigma_+)^2]/
	       [(\sigma_+)^2(2-\exp        \nonumber
\\ &&(-(\rho_c)^2/2(\sigma_-)^2))] \;   for \;   \rho_m<\rho
\label{equationc}
\end{eqnarray}

The values of $\sigma_-$ ($\sigma_+$)controls the width of the low (high)
density branch. Here, $\rho_c$ is the end
local density of the hole-poor branch. $\rho_m$ is
the starting local density of the hole-rich or
metallic branch. Both $\rho_c$ and $\rho_m$ are shown in Fig.1a for
the $\langle \rho \rangle =0.16$ case.
We can get a reliable estimation of the $ \sigma_+$ values from
the experimental STM/S Gaussian histogram  distribution
for the local gap\cite{Pan} on an  optimally doped
$Bi_2Sr_2CaCu_2O_{8+x}$\cite{Mello01}.

According to the percolation theory, the site percolation
threshold occurs in a square lattice
when 59\% of the sites  are filled\cite{Stauffer}.
Thus, we find the density where the hole-rich branch percolates
integrating $\int P(\rho) d\rho$ from  $\rho_m$ till the integral
reaches the value of 0.59 where we define $\rho_p$.

%\section{The Phase Diagram }
%  

There are several different approaches  which can be used to obtain
$T^*(\rho(r))$. Strictly speaking, due to  the non-uniform charge
distribution, we do not have translational symmetry and
we should use a method which takes the disorder into
account. 
Since our purpose here is to demonstrate the
feasibility of the percolating scenario,we will take the
simplest way, that is, we will follow a BCS-like approach with an extended
Hubbard Hamiltonian to derive a curve for the temperature onset of vanishing
gap  as function of the local hole concentration $\rho$. 
Using the experimental parameters for the dispersion relation
and hopping integrals derived from experiments and band calculations
appropriate to the LSSCO family\cite{Mello99b}, we derived 
the onset of vanishing
gap as function of the doping level. This curve is taking  as the
$T^*(\langle \rho \rangle)$ boundary as shown in fig.1b.

%\subsection{Subsection}

\begin{figure}[!ht]
\begin{center}
\includegraphics[width=0.6\textwidth]{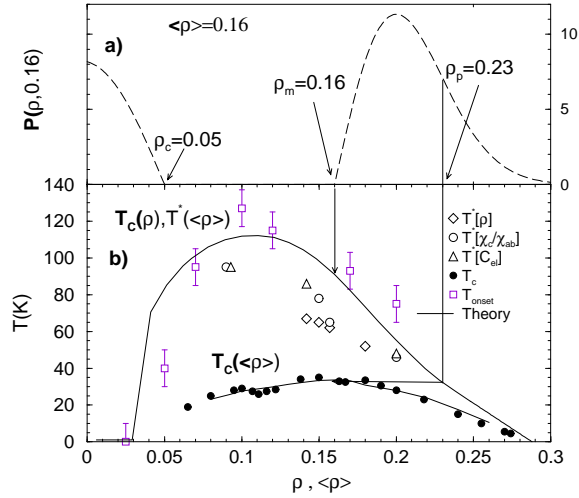}
\end{center}
\caption{Phase diagram for the LSCO family. To explain how
$T^*(\langle \rho \rangle)$ and $T_c(\langle \rho \rangle)$  are obtained,
we plot in (a) the probability distribution for the optimal compound
with $\langle \rho \rangle=0.16$, P$(\rho,0.16)$. The arrows shows  $T^*(0.16)$
and the  percolation threshold at  $\rho_p=0.23$ with
$T_c(\langle \rho \rangle=0.16)= T_c(0.23)$.
The  experimental data are taken from Ref.~\cite{Oda}
and $T_{onset}$ is taken from the flux flow experiment
(open squares)\cite{Xu}.}
\label{samplefig}
\end{figure}

The theoretical curves in Fig.1 are derived in the follow way:
For each local density $\rho > \rho_m$, the
maximum value of $T^*(\rho)$ is equal $T^*(\rho_m)$.
This is the onset temperature of the superconducting gap 
and therefore $T^*_c(\rho_m)=T^*(\langle \rho \rangle)$.
$T_c(\langle \rho \rangle)$
is estimated in the following manner: we
calculate the  maximum temperature in which the superconducting
region  percolates
in the metallic branch. The percolation occurs when all the clusters with
local density between $\rho_m$ and $\rho_p$ are superconducting
as shown for $\langle \rho \rangle=0.16$. 
The value of $\rho_p$ can be seen in the panel
following the arrow which shows that  $T^*(\rho_p=0.23)$ is equal the
superconducting critical temperature of the compound,
$T_c(\langle 0.16\rangle)$. 

%\section{Discussion}

There are several HTSC phenomena  which are not well understood
and  can be  explained with
the percolating ideas derived from our model and
calculations. Here we name just a few:\\
i) The steady decrease of the zero temperature gap $\Delta_0(\langle 
\rho \rangle)$ with the
doping $\langle \rho \rangle$. ii) The resitivity 
deviation from the linear behavior at $T^*$. iii) The existence of
superconducting clusters between $T^*$ and  $T_c$ easily
explains the appearance of local diamagnetic or Meissner domains\cite{IYS}
and the Nerst flux flow effect above $T_c$\cite{Xu}.
iv) The suppression in the specific heat $\gamma$ term found
in underdoped
compounds of different families.

In summary, we have proposed a quantitative  novel and general  approach
to the phase diagram of the
high-$T_c$ cuprates superconductors in which the pseudogap is
the largest superconducting gap among the  superconducting regions in an
inhomogeneous compound. The critical temperature $T_c$ is
the maximum temperature for which these superconducting regions
percolates. This method is  also suitable  to provide
insights in the normal phase properties.

\end{document}